\documentclass[a4paper,12pt]{article}

\usepackage{aas_macros}
\usepackage{setspace}
\usepackage{graphicx}

\newcommand\cS{{\cal S}}

\catcode`@=11

\newcommand{\Set}[1]{\left\{{#1}\right\}}       
\newcommand{\Event}[1]{\left\{\mskip 0.25\thinmuskip#1\mskip 0.5\thinmuskip\right\}}

\def\checkEventSub#1_#2%
{%
  \@ifnextchar^{\checkEventSubSup{#1}_{#2}}{\checkEventOpen{{#1}_{#2}}}%
}%

\def\checkEventSup#1^#2%
{%
  \@ifnextchar_{\checkEventSupSub{#1}^{#2}}{\checkEventOpen{{#1}^{#2}}}%
}%

\def\checkEventSubSup#1_#2^#3{\checkEventOpen{{#1}_{#2}^{#3}}}
\def\checkEventSupSub#1^#2_#3{\checkEventOpen{{#1}^{#2}_{#3}}}

\newcommand{\checkEventOpen}[1]%
{%
  \@ifnextchar\bgroup{{#1}\!\Event}{%
    \ifx_\@let@token
      \checkEventSub{#1}%
    \else\ifx^\@let@token
      \checkEventSup{#1}%
    \else\ifx(\@let@token
      {#1}%
    \else\ifx\@let@token\empty
      {#1}%\mskip 0.5\thinmuskip
    \else
      {#1}%\mskip -0.5\thinmuskip
    \fi\fi\fi\fi}%
}%                                       % for emacs }-matching

\renewcommand{\checkEventOpen}[1]%
{%
  \@ifnextchar\bgroup{\def\tmpABC{{#1}\!\Event}\tmpABC}{%
    \ifx_\@let@token
      \def\tmpABC{\checkEventSub{#1}}%
    \else\ifx^\@let@token
      \def\tmpABC{\checkEventSup{#1}}%
    \else\ifx(\@let@token
      \def\tmpABC{{#1}}%
    \else\ifx\@let@token\empty
      \def\tmpABC{{#1}}%\mskip 0.5\thinmuskip
    \else
      \def\tmpABC{{#1}}%\mskip -0.5\thinmuskip
    \fi\fi\fi\fi\tmpABC}%
}%                                       % for emacs }-matching

\renewcommand\d{\@ifnextchar.{{\rm d}}{\,{\rm d}}} 

\catcode`@=12

\usepackage{fullpage}

\usepackage{natbib,natbibspacing}
\title{Straight to the Source: Detecting Aggregate Objects in Astronomical Images with Proper Error Control}
\author{David A. Friedenberg and Christopher R. Genovese}
\date{}
\begin{document}

\maketitle

\begin{abstract}
The next generation of telescopes will acquire terabytes of image data on a nightly basis.
Collectively, these large images will contain billions of interesting objects,
which astronomers call \textit{sources}.
The astronomers' task is to construct a catalog detailing the coordinates and other properties of the sources. 
The source catalog is the primary data product for most telescopes
and is an important input for testing new astrophysical theories,
but to construct the catalog one must first detect the sources.
Existing algorithms for catalog creation are effective at detecting sources,
but do not have rigorous statistical error control. 
At the same time, there are several multiple testing procedures that provide rigorous error control,
but they are not designed to detect sources that are aggregated over several pixels.
In this paper, we propose a technique that does both, by providing rigorous statistical error control 
on the aggregate objects themselves rather than the pixels.
We demonstrate the effectiveness of this approach on data from the Chandra X-ray Observatory Satellite.
Our technique effectively controls the rate of false sources, yet 
still detects almost all of the sources detected by procedures that do not have such rigorous error control and have the advantage of additional data in the form of follow up observations, which will not be available for upcoming large telescopes.
In fact, we even detect a new source that was missed by previous studies.
The statistical methods developed in this paper can be extended to problems beyond Astronomy,
as we will illustrate with an example from Neuroimaging.
\end{abstract}

David Friedenberg is Doctoral Student, Department of Statistics,
Carnegie Mellon University,
Pittsburgh PA 15213 (E-mail:dfrieden@stat.cmu.edu).
Christopher Genovese is Professor, Department of Statistics,
Carnegie Mellon University,
Pittsburgh PA 15213 (E-mail:genovese@stat.cmu.edu).
This work is supported by National Science Foundation (NSF) grants 486560 and DMS0806009, National Institute of Health (NIH) grant 1R01NS047493, National Aeronautics and Space Administration (NASA) grant NNX07AH61G and the Bruce and Astrid McWilliams Center for Cosmology.
The authors would like to thank Arthur Kosowsky and Neelima Seghal for the Atacama Cosmology Telescope data, Peter Freeman for his help with the Chandra X-ray telescope data, David Heeger and Eli Merriam for the fMRI data, and Rebecca Nugent, Chad Schafer, and Larry Wasserman for their helpful discussions.

\newpage

\section{Introduction}

The typical astronomical image records the intensity of light, over some range of frequencies,
across a section of sky that contains many celestial objects
of various size, shape, and luminosity.
The image's pixels correspond to an array of light-sensitive detectors in the telescope, % -- as in a digital camera
and each pixel essentially counts how many photons have
struck the corresponding detector during the exposure.
But the photons recorded in the image do not come solely from the objects of interest, or \emph{sources};
thermal noise and background emissions (collectively called \emph{background}) corrupt the data and obscure the signature of the objects.
Moreover, diffraction and atmospheric effects blur the image, reducing resolution and washing
out the fainter signals.
In the (astronomical) source detection problem, one is given such an image
and seeks to construct a \emph{catalog}
that gives the coordinates (and often other properties) of sources in the image.

A source catalog is the basic data product of most astronomical surveys
and the basic input to the scientific process.
This has been true for some time.
Early catalogs -- from
the data of ancient astronomers Shi Shen and Hipparchus, each cataloging about 1000 stars,
to the compendium of deep-sky objects produced by William and Caroline Herschel in the 1700s \citep{catalogue-of-nebulae} %\emph{Catalogue of Nebulae}
-- were based on direct visual observations.
Later work, especially in the 20th century, used photographic plates,
both improving resolution and allowing the detection of much fainter objects.
But either way, compiling a source catalog would be a slow and painstaking affair,
often requiring years to collect data on only a handful of objects.
Until recently, catalogs comprising a few hundred objects were large, a few thousand were epic.

All this changed with the advent of new technologies -- digital imaging, advanced designs for telescope mirrors,
and computer automation -- and with increases in available computing power and storage.
With relative suddenness, astronomers found that they could observe wider, deeper, and faster than ever before.
They could sweep the sky searching automatically for objects of a certain type, 
they could collect data on many objects in parallel, and they could observe
a multitude of faint objects that would previously have gone undetected.
The Sloan Digital Sky Survey \citep{sdss} has measured hundreds of millions of objects.
The upcoming Large Synoptic Survey Telescope (LSST, \citep{lsst})
will scan the entire sky every few days, collecting several terabytes of data per night
into a catalog comprising \emph{billions} of objects.
Over the past two decades, astronomy has gone from data poor to data rich.

And therein lies both opportunity and challenge.
The opportunity lies in the richness and importance of the scientific questions that these massive data sets can answer.
The challenge lies in the sheer scale of the data analysis.
While as a general rule in science, more data is better,
there is a reason that astronomers often describe the coming bounty of data in quasi-biblical terms
-- a flood, a tsunami, an onslaught.
The next generation of astronomical catalogs, including the LSST,
will be so large that even simple operations -- such as a basic query of
the entire catalog -- will be computationally prohibitive,
and yes that does account for Moore's law.

This influx of data has motivated several statistical innovations in the field of Astronomy leading to the emergence of the sub-field, astrostatistics. New statistical innovations have had a significant impact on several important and cutting edge Astronomy problems, for a small sampling see \citet{vandyk-2009-3}, \citet{loredo}, \citet{rice}, \citet{npicmb}, and \citet{richards}.

Besides the massive size of the data, another issue is controlling the rate of errors in the catalog.
In the past, where every object in the catalog was observed manually,
sources were often missed, located incorrectly, or created spuriously.
Follow-up observations can often be made for all or most of the sources,
reducing false positive and false negative identifications to a manageable level.
But in the near future, the sheer number of objects in the catalog
will preclude comprehensive follow-up observations by human beings.
Scientific studies of the catalog will likely need to be based on
samples or selections made by automatic, statistical criteria.
But no matter how carefully these criteria are constructed,
there will be objects that are misclassified.
To use the resulting samples effectively for scientific inference,
it will be necessary for the method to provide tunable control
of error rates.
Thus, new statistical and computational methods will be needed to construct and analyze the next generation
of astronomical source catalogs.

In this paper, we develop a multiple-testing-based method for the source detection problem
that has several advantages over existing techniques, especially for the analysis of large-scale surveys like the LSST.
Although we discuss our method in the context of astronomical source detection, 
the method applies to a wide range of similar problems such as neuroimaging \citep[e.g.][]{neuro} and remote sensing \citep[e.g.][]{remote},
and we give such an example in a later section.

We assume that the input image is an $n\times m$ array of pixels, with the value recorded
at pixel $(i,j)$ denoted by $Y_{ij}$.
The photons that contribute to $Y_{ij}$ arise from two components: \emph{sources}, 
the emissions produced by the celestial objects of interest, and \emph{background}, 
which includes thermal noise, the emissions of unresolved objects, interfering radiation sources,
atmospheric emissions, and all other anomalies or artifacts.
astronomical images essentially measure photon counts
for which a Poisson model is appropriate \citep{poissModel}.
So for our base model, we assume that the $Y_{ij}$'s are independent with
\begin{equation} \label{eq::simple-poisson-model}
Y_{ij} \;{\rm distributed\ as\ } {\rm Poisson}\langle \lambda_{1,ij} + \lambda_{0,ij}\rangle,
\end{equation}
where $\lambda_{1,ij} \ge 0$, $\lambda_{0,ij} \ge 0$ denote the mean intensity of
sources and background, respectively and
$\lambda_{1,ij}+\lambda_{0,ij} > 0$.  The idea here is that the
pixels are measuring the counts in disjoint cells of a Poisson random
field across the sky.  This applies to good approximation for
space-based observations like those reported in Section 2.  For
ground-based observation, the image is in addition blurred by
atmospheric turbulence, so the
$Y_{ij}$'s are no longer strictly independent.  However, the Poisson
model in Equation \ref{eq::simple-poisson-model} still holds to
reasonable approximation.  In data sets where the counts are high, the
Poisson random field can be further approximated by a Gaussian random
field.  In this paper, we utilize a technique originally developed for
the Gaussian model and generalize it so that we can accommodate a wide
variety of models.

\begin{figure}[h!] 
        \centerline{\includegraphics[scale=.35]{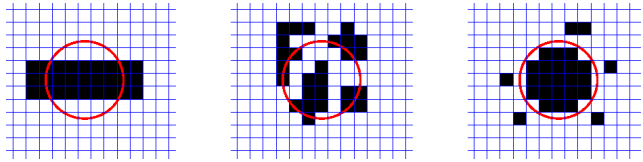}}
        \caption{The 45 pixels that have any overlap with the red circle are considered sources and black pixels indicate sources detected via some detection algorithm. Each of the three images have 6 Type I error pixels and 24 Type II error pixels. The detection in the left image captures the center of the source but clearly misses the shape. The detections in the center image show several different sources that have some overlap with the true source. The detection in the right image captures the center of the true source and has some spurious noise detections. While all these pictures have the same number of pixel-wise Type I and Type II errors they lead to very different conclusions about the number, shape and location of the sources in the image, thus a testing criteria based on the aggregate sources, instead of pixels, is necessary.   }
        \label{fig:type1and2}
\end{figure}

The source detection problem 
is to identify which pixels contains sources
and thus to separate the sources from the background.
If $\lambda_{1,ij} > 0$, then we take pixel $(i,j)$ to be a source pixel; otherwise, it is a background pixel.
So it is natural to consider this as a multiple testing problem
with the null hypothesis at each pixel being that $\lambda_{1,ij} = 0$.
At a coarse level, we want to characterize the set $\cS = \Set{(i,j):\; \lambda_{1,ij} > 0}$ of source pixels,
but our more specific goal is to identify and locate the underlying sources,
so that an accurate catalog can be constructed.
This requires a more stringent criterion for success
because the objects are coherent, localized aggregates.
As Figure \ref{fig:type1and2} shows, with the same number of pixel-wise type I and type II errors,
it is possible to get widely varying accuracy in the resulting catalog.
%Correctly identify a large number of source pixels but getting the shape wrong
%will produce a grossly inadequate catalog.
Put another way, our loss function operates on the catalog, not the pixels themselves.
%ATTN:SMITH

In this paper, we extend the False Cluster Proportion (FCP) controlling procedures
introduced by Perone Pacifico et al. (2004) to make it
effective for controlling the rate of false sources detected in astronomical images.
As we will show below, the original FCP procedure does not perform well with
the Poisson statistics common in the source detection problem and even where
the Gaussian assumption holds, does not yield sufficient power
to be viable for the astronomical source detection problem. We generalize
the technique so that it applies to a wider range of noise models. We also
introduce a new transform, which we call the \emph{Multi-scale Derivative},
that enhances sources and significantly improves power. Taken together,
these extensions lead to a new procedure that we call the
Multi-scale False Cluster Proportion (MSFCP) procedure. This gives a powerful
source detection technique that provides rigorous control over the
rate of false \emph{sources}, where techniques in current use provide control
over the rate of false \emph{pixels}, if they provide any control at all.
We demonstrate both the 
excellent power and error control on a very deep and high resolution
telescope image from the Chandra X-Ray Observatory. MSFCP has
detection power competitive with the existing procedures, even
detecting a source that was overlooked in previous studies while at
the same time maintaining rigorous control over the rate of false
sources. Although the source detection problem is ubiquitous in Astronomy
it also occurs in other settings and we present an example of how FCP
concepts can be extended to detecting bands of neural activity in the brain.

Due to its prevalence and difficulty, there have been a variety of approaches to the source detection problem,
both in the statistical and astronomical literature.
Given the recent explosion in research on multiple testing,
there is a plethora of available techniques that can be applied directly to the
pixel-wise hypothesis tests to reconstruct $\cS$,
including \citet{bh}[BH], \citet{stepup}, \citet{storey}, \citet{suncai}, and \citet{rice} .
However, because these methods give error-rates in terms of the individual pixel-wise tests,
it is not obvious how to translate these error-rates to make inferences about the underlying sources.
Multiple testing approaches that are less pixel-centered have been developed
in the related problem of analyzing functional magnetic resonance imaging (fMRI) data.
In this problem, the sources are regions of neural activity that reveal themselves
through a measurable change in blood flow response.
\cite{wor96} and \citet{wor02} use level sets of a random field formed from test statistics to identify regions containing sources.
\citet{cba} construct a test based on clusters instead of pixels in the fMRI setting,
but this technique takes advantage of a temporal dimension that is not available in many general (e.g., astronomical) problems.

%SOMETHING ABOUT THE ASTRONOMERS' OUTLOOK AND ERROR RATES.
%PRE-PROCESSING/FILTERING, PIPELINE
%CONTEXT FOR ASTRONOMICAL USE
Source detection understandably garners much attention in the astronomical literature.
Astronomers address source detection as one step in a data-processing pipeline --
the series of operations performed on the data from collection until catalog.
These include, but are not limited to,
corrections for atmospheric effects, image registration, filtering out unwanted signals, as well as source detection.
These pipelines are typically planned and developed well before the instrument is operational.
During the planning stage, simulations are run to test the pipeline including the source detection algorithm \citep[e.g.,][]{sehgal}.
Astronomers typically do not require their algorithms to satisfy any formal performance criteria;
rather, they apply an empirical criterion,
using data simulated to look like a real telescope image to calibrate the error rate for detected sources that will be expected in practice.
Of course, this depends on the simulated and real data being both quantitatively and qualitatively similar.
While great effort and ingenuity are applied in constructing realistic simulations -- sometimes years of computing time
for a single run -- the simulations still rely on untested and unstated assumptions that may fail when
the instrument comes on line.

The source detection methods used by astronomers fall into three general classes: 
simple thresholding, peak-finding algorithms, and Bayesian algorithms.
Simple thresholding consists of choosing an intensity cutoff for pixel-wise statistics and classifying any pixel above threshold as a source.
It is popular because it simple and fast,
easily computed by the popular SExtractor software \citep{sext}.
Before thresholding, filters are often applied to the raw data
to suppress confounding background signals.
In \citet{vik} and \citet{melin}, 
Matched Filters are used to isolate the signal,
and then thresholds are determined using simulated data to create catalogs in X-ray and Radio telescope images respectively.
\citet{2002AJ....123.1086H} also use simple thresholding, choosing the threshold to control the False Discovery Rate via the method of \citet{bh},
but they use simulated telescope images to calibrate 
between rate of false pixels and the rate of false sources.

Peak-finding algorithms search for local maxima in the denoised image and catalog them as sources. 
\citet{vale} and \citet{sehgal} use peak-finding algorithms to look for large galaxy clusters in simulated radio telescope images.
The wavelet-based technique of \citet{peter} is commonly used to detect X-ray sources as in \citet{valt} and \citet{1msec}.
\citet{Damiani} and \citet{gonzaleznuevo-2006-369} provide a good overview
of the popular Mexican Hat wavelet as a tool for source detection. 
Several implementations exist in software and
most are specifically tailored for a certain instrument or type of problem.
Pixel-wise error rates for wavelet source detection can sometimes be determined analytically
but are more often approximated from simulations.
As with simple thresholding,
simulations are often used to indirectly estimate the rate of false sources,
the error rate of interest, from the rate of false pixels.
%ATTN connect to above and contrast as needed

Bayesian techniques have become popular with astronomers and several have applied Bayesian methods to source detection as in 
\citet{2007ApJ...661.1339S}, \citet{strong}, and \citet{2003MNRAS.338..765H}. 
Bayesian detection algorithms typically define models for sources, 
often two-dimensional Gaussians,
and attempt to distinguish them from backgrounds via inference on a posterior distribution.
\citet{gug} use Bayesian mixture models to separate sources from background,
while \citet{fastBayes} propose ways to speed up Bayesian source detection,
which can be computationally slow for large problems.
Bayesian algorithms typically require more assumptions to be made a priori
and can be more computationally expensive than thresholding or peak-finding algorithms.

%ATTN check harshness, somewhat awk -- more concrete
%arxiv cite?

%ATTN:SMITH BEGIN
Our goal in source detection is to detect the relevant objects, not pixels,
while controlling the error rates.
Statisticians have developed numerous methods for dealing with error rates
but have not been focused on detecting aggregate objects,
while astronomers have been thinking about detecting objects,
but without a rigorous approach to controlling error rates.
We propose a technique that does both: control the error rates and make our inference about the sources themselves.

We demonstrate our techniques on an important data set from the Chandra X-ray Observatory,
one of the most powerful telescopes in existence.
For the Chandra data, 
our goal is to detect the X-ray sources with a bound on the error rate for sources.
We describe an approach due to \citet{Pacifico04} that gives us a probabilistic bound on the error rate for sources, 
and apply it to the Chandra data.
We find that while this approach gives us the error control we want,
it does not have good power when compared to other techniques.
We introduce a generalization of the technique
that allows for us to keep a probabilistic bound on the error rate for sources under more general conditions.
We then introduce a new Multi-scale technique that is designed to enhance sources and thus increase power.
We then integrate it with our generalized procedure to get the MSFCP procedure
which increases power while maintaining control over the error rate.
This improvement is evident when we revisit the Chandra data --
we show that our power using MSFCP is competitive to two algorithms commonly used by astronomers,
but with superior error control.
Furthermore, using our procedure we detect a X-ray source which had gone undetected in the original analysis of the data by astronomers.
We then provide a brief description of how these techniques can be used outside the realm of Astronomy with an application to high-resolution neuroimaging data.
We conclude with an overview of our results and directions for further study.
%ATTN:this is the most generic sentence ever
%ATTN:SMITH END

\section{The Data}
We demonstrate our techniques on data from the Chandra X-ray observatory
\citep{chandra}, one of the most powerful X-ray telescopes in the world.
Chandra orbits Earth, at approximately one
third the distance to the Moon. Being outside of the Earth's
atmosphere allows for extended observing time without atmospheric disruption of
the X-ray signals themselves.
X-rays are emitted when matter is heated to millions of degrees, such as in the hot gas
surrounding large galaxy clusters, during supernovae, or when matter circles a
black hole. Resolving discrete sources of X-rays in space has been an important
problem in Astronomy for several decades. Because studying X-ray emissions from
galaxy clusters can answer many of the fundamental questions in Astronomy and Cosmology, several generations of instruments have been developed specifically
for this task. From X-ray data, astronomers can infer the evolution of galaxies,
which informs us about the evolution of the universe. For instance, recent X-ray
observations of the Bullet Cluster have provided the most compelling evidence of
dark matter \citep{bullet}, a mysterious form of matter that does not strongly
interact with visible matter but is postulated to account for approximately 22\%
of the mass of the universe.

\begin{figure}[h!] 
        \centerline{\includegraphics[scale=.35,angle=90]{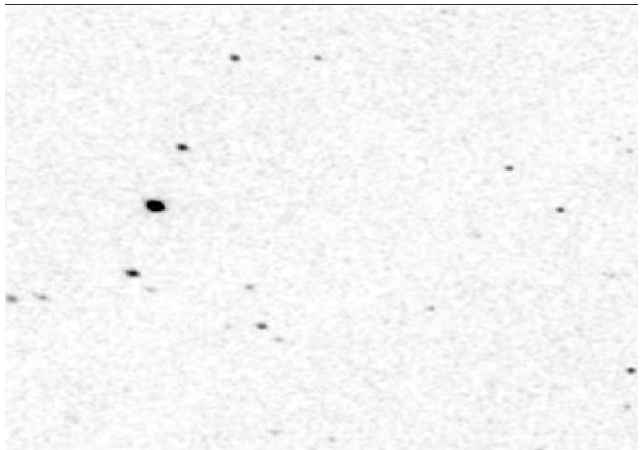}}
        \caption{A 512 pixel by 512 pixel patch of the Chandra Deep Field South.
The image was smoothed and then all pixels with count greater than 15 were set to 15. This image is the log of that image plus one. The transformations were performed so that the brightest sources do not wash out the fainter ones for display purposes. They were not used in any of the analysis. The complete data set is publicly available
from http://cxc.cfa.harvard.edu/cda/}
        \label{fig:xray}
\end{figure}

The data used in this paper comprise part of the Chandra Deep Field South (CDFS),
a composite image of 11 Chandra observations of a small patch of sky with
nearly one million seconds of observing time. This is a long exposure which
means we should be able to resolve very distant and faint sources. This piece of
sky was selected because it had low interference from our own galaxy and no
bright stars in its vicinity. The CDFS has area approximately one half the angular size
of the moon. It is also in a location that can be observed by several
complementary ground-based telescopes. The part of the CDFS that we will analyze
and discuss in this paper can be seen in Figure \ref{fig:xray}. The first step
in converting the raw
image data into an actionable scientific dataset is cataloging all the sources
in the image. The sources in the CDFS are thought to be active galaxies and
quasars with massive black holes in their centers. We will provide a new
analysis of this dataset that not only detects these important X-ray sources,
but also makes a probabilistic guarantee on the rate of falsely detected sources.

The original analysis of the CDFS was published in \citet{1msec} [GI]. Their main
catalog was created by combining the catalogs from two different source
detection algorithms: a modified version of the SExtractor algorithm
\citep{sext}, and the WAVDETECT algorithm \citep{peter}. The SExtractor algorithm
estimates the background and catalogs as a source any
region that is $2.4\sigma$ above the background that is also at least 5 pixels
in area. These parameters are determined using images simulated to look like the
real telescope images. They are selected to create a large catalog which will
likely include several false sources. Independently, the WAVDETECT procedure was
run using a wavelet transformation specifically tuned for this type of
image to detect sources. Again, parameters are calibrated using simulated images.
The two catalogs are merged and then refined using follow-up observations in an
effort to remove sources that may be false. In this step they are attempting to
replicate the detection with independent observations, thus if an object is in
the catalog we are fairly certain it is real since it has been observed multiple
times, often with a completely different instrument. It is important to note that this detection strategy 
is designed for a small patch of sky with the ability to follow-up all potential
detections. In this scenario, it makes sense to use a detection scheme that
casts a wide net. Since we can go back and verify or reject each potential
detection we will tolerate a larger number of spurious sources initially in
exchange for the ability to detect some fainter sources. This strategy will fail
with newer telescopes that will be scanning large areas of sky and making so
many detections that it will be impractical to follow-up each and every
detection. Instead we want a strategy that will automatically detect real
sources reliably while controlling the rate of false ones in one pass through
the data.

\section{False Cluster Proportion Algorithm}
A reasonable first approach for detection is to conduct a hypothesis test at
each pixel individually.  We want to test whether a pixel is a background pixel
(null) or a source pixel (alternative). After conducting the appropriate pixel-wise test, a multiple
testing correction can be applied such as the BH method. A problem
with the pixel-wise testing approach is that the unit of inference are individual pixels. A
pixel is an artificial unit related to the resolution of instrument. What we are
interested in are objects composed of collections of adjacent pixels. Therefore,
we want the unit of inference to be a cluster as opposed to a pixel. An alternative to pixel-wise testing is the False Cluster Proportion (FCP) procedure introduced by \citet{Pacifico04}. 
The method is designed to bound the rate of false regions of a random field. FCP treats the image as a realization of an underlying
random field and then derives a confidence superset for the location of the true
nulls (background regions). Denote the the unknown
true nulls as the set $S_0$ and the confidence superset as $U$. $U$ is a $1-\alpha$ confidence superset for $S_0$ if 
\begin{equation}
 \label{superset}
 P(U \supseteq S_0)\geq 1-\alpha
\end{equation}
Let $L_t$ denote the level set of
all pixels that have intensity greater than t. $L_t$ can be decomposed into
its connected components. Every pixel in $L_t$ is grouped with all its neighboring
pixels that are also in $L_t$ creating clusters of pixels with intensity greater
than t.  This decomposition yields a set $C_t$ of $k_t$ clusters
$C_t=\{C_{1,t},C_{2,t},...C_{k_t,t}\}$. For any cluster $C$ the cluster is declared
false if
\begin{equation}
\frac{\lambda(C \cap S_0)}{\lambda(C)} \geq \epsilon \label{crit}
\end{equation}
In a pixelized image one can use the counting measure and $\epsilon$ is a pre-specified
tolerance parameter.  The goal is to come up with a threshold $t$
that insures the proportion of detected objects that are false is
sufficiently low. Define the true false cluster
proportion, $\Xi(t)$:

\begin{equation}
 \Xi(t) =  \frac{\#\{1\leq i \leq k_t : \frac{\lambda(C_{it}\cap
S_0)}{\lambda(C_{it})}\geq \epsilon\}}{k_t}
\end{equation}

This quantity can be bounded by calculating the false cluster proportion
envelope $\bar{\Xi}(t)$, where

\begin{equation}
\bar{\Xi}(t) = \frac{\#\{1\leq i \leq k_t : \frac{\lambda(C_{it}\cap
U)}{\lambda(C_{it})}\geq \epsilon\}}{k_t}
\end{equation}

Then from \citet{Pacifico04}
\begin{equation}
P(\Xi(t) \leq \bar{\Xi}(t)~ \forall t) \geq 1 - \alpha
\label{guar}
\end{equation}

Suppose we have a confidence superset $U$, next we need to find the
value $t_c$ such that $\bar{\Xi}(t_c) = c$ where $c$ is the false
cluster proportion value which we do not wish to exceed. To do this
one must perform a search over the possible values of t. Once $t_c$
has been determined, take the $k_{t_c}$ clusters from the level set
$L_{t_c}$ as the detected objects. They have the property that with
probability $1-\alpha$ the proportion of the detected objects which
are false detections are less than or equal to $c$.

To use FCP one needs a confidence superset $U$, which satisfies Equation
\ref{superset}.
\citeauthor{Pacifico04} calculate $U$ by looking for the smallest set $A$ for which the null of the following test cannot be rejected.
\begin{equation}
\label{supersettest}
H_0:A\subset S_0 \mbox{ versus } H_1:A \not\subset S_0
\end{equation}
In practice, the maximum pixel value in the set $A$ is used as the test statistic.
The p-value is then

\begin{equation}
\label{pval}
 p(x,A)=P_0(\max (Y_{ij} \in A) \geq x)
\end{equation}

$U$ is calculated by starting with the entire image and then progressively
removing the pixels with the highest intensity (those least likely to be part of
the background) until the p-value of the remaining set passes $\alpha$.
The p-value in Equation
\ref{pval} is calculated using the Piterbarg approximation \citep{pite}.
The Piterbarg approximation assumes that the
field is a locally stationary Gaussian random field with quadratic covariance.
If  X(s) is a homogeneous Gaussian random field then $\frac{X}{\sigma}$ is
locally stationary with quadratic covariance if for some matrix B,

\begin{equation}
\rho (s) = 1 -s^T B s + o(||s||^2) \label{quad}
\end{equation}

This method yields a valid confidence superset $U$ for the case of Gaussian Random Fields that satisfy Equation \ref{quad},
which can then be incorporated into FCP to find the appropriate value
for $t_c$.
The clusters detected using the cutoff $t_c$  will, with high
probability, have a false cluster rate less than the bound $c$.
Henceforth we will refer to this approach as the PP method; for full
details see \citet{Pacifico04}.

Once we have run FCP, astronomers can plug the value $t_c$ as an argument to the SExtractor software
and generate the catalog without having to learn any new software and at
computational speed to which they are accustomed.  Many astronomers use SExtractor for
source detection either using the default settings or by ``playing with various
parameter combinations and inspecting the results by eye'' \citep{spitz}. FCP
gives astronomers a more principled way to choose the parameters in the software
they are already using.

\section{FCP Applied to CDFS}
\begin{figure}[ht] 
        \centerline{\includegraphics[scale=.4]{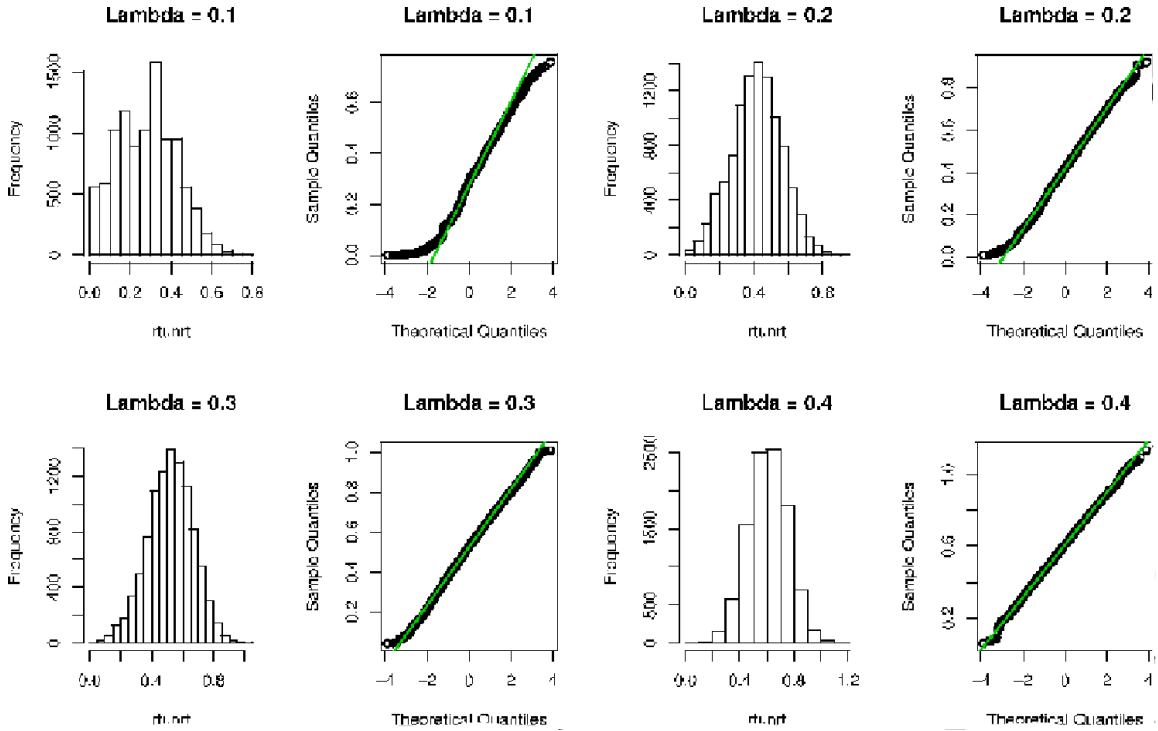}}
        \caption{Simulation of a 100x100 image of Poisson Background which is
then smoothed with a Gaussian, with standard deviation of 1 pixel. The square
root is then taken at each pixel. The histogram shows the pixels become
approximately normal for values of lambda above .2. The Quantile-Quantile plots
show the data compared to a perfectly normal distribution. The closer the points
lie to the line y=x (green), the closer the data is to a normal distribution}
        \label{fig:poiss}
\end{figure}

In the CDFS data, the objects we are looking for are active galaxies and quasars
which are emitting X-rays. We would like to detect as many of these objects as
possible without making spurious detections, that is classifying what is really
just noise as an X-ray emitting object. 
Each pixel records the number of photons that strike the detector and are modeled as Poisson random variables.
The mean background in these images is small, approximately 0.3, which violates the assumption
of the PP procedure which assumes a Gaussian background. 
Our initial approach to dealing with the non-gaussianity is to filter the data
so that the Gaussianity assumption is more appropriate.
Specifically, we
first smooth the data and then apply a pixel-wise transformation.
Smoothing the data with a Gaussian filter effectively aggregates
counts so we preserve most of the structure of the Poisson data even
though they are no longer counts. There are also many more unique values which
makes it closer to high rate Poisson data than low rate Poisson data, which is mostly zeros
and ones. We then take the square root transformation which makes the background
approximately Normal. The square root transformation has been shown to normalize
high count Poisson data \citep{anscombe}. Using simulations, we have tested the effectiveness of
our smooth-then-square-root procedure for low rate Poisson data and found
that for rates as low as 0.2, it yields approximately normal
data. Examples for different rate parameters is shown in Figure \ref{fig:poiss}.

\begin{figure}[h!tp] 
        \centerline{\includegraphics[scale=.35,angle=90]{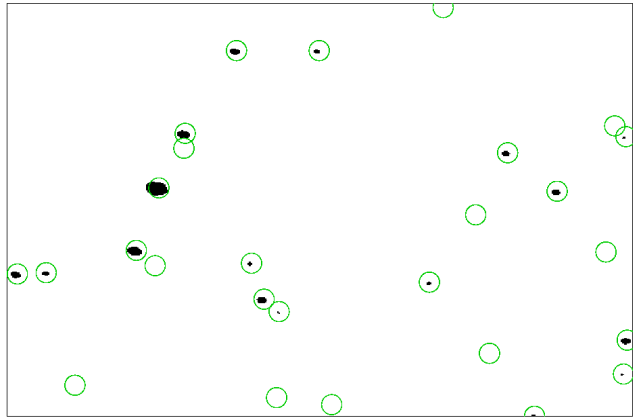}}
        \caption{CDFS data after applying the False Cluster Proportion procedure
using the PP method to calculate U. Black indicates sources detected by FCP
and the circles are detections from both the Main Catalog and Secondary Catalogs
from GI. There are 17 detections from FCP all of which are in the
Catalog and 9 missed sources.}
        \label{fig:FCPVanila}
\end{figure}

After the normalizing transformation, 
we perform a standard Z-test at each pixel, 
assuming a constant background mean,
and calculate the 95\% confidence superset for the background by applying the PP method to the test statistics.
We then run FCP on the test statistics as described in Section 2, bounding the rate of false sources to be less than 10\%. Because our $U$ is very conservative (it contains every pixel from the null with probability .95) we select $\epsilon = .99$ to maximize our detection power. 
Nothing in our procedure relies on information about the
telescope or knowing the nature or distribution of the sources themselves. We
only assume that our normalizing transformation does indeed give us a background
that is approximately Gaussian, an assumption which can easily be checked.
FCP, using the PP method to calculate $U$, detects 17 sources, all of which are in
the previously published GI catalog as seen in Figure \ref{fig:FCPVanila}. Since
this catalog has been verified with follow up observations we
believe all the detections are real. However, this analysis is clearly missing many clusters
that should be detectable -- nine of the 26 clusters in the Main catalog of
GI were missed using FCP with the PP method. 

One way to boost the power is to run the data through an appropriate filter that
will enhance the sources and help separate them from the background. However, to
do that and maintain the desired control over the rate of false sources, we need
to adapt FCP so that it can handle more varied noise conditions. 
This means moving away from the PP method to a more general procedure for calculating $U$. 
In the following section we propose a procedure for calculating $U$ under any noise
distribution that we can simulate. This will allow us to run FCP on filtered
data and achieve higher power while maintaining control over the rate of false
sources.

\section{Generalizing FCP}
The False Cluster Proportion Algorithm requires the derivation of a
confidence superset for the background of an image.  What we have been
referring to as the PP method, which is based on the Piterbarg
approximation, is limited to cases where test statistics are or can
be transformed to a Gaussian Random Field that satisfies Equation
\ref{quad}.  But in many situations the data are not Gaussian and
finding an acceptable transformation is infeasible.  For example,
filtering techniques commonly used to enhance sources will usually not
satisfy the assumptions of the PP method.

To broaden the scope of the technique, we have developed a simulation
procedure that produces an accurate $1-\alpha$ confidence superset for
any noise distribution from which we can sample numerically.  Assume
our data are a rectangular image with $N$ total pixels and known noise
distribution F.  We simulate noise images and calculate the maxima of
subsets of the images.  We build up empirical distributions for these
maxima and use them to calculate the p-values which, in the PP method,
were calculated using the Piterbarg approximation.  We can then plug
the p-values into the FCP procedure and get the False Cluster
Proportion guarantee (Equation \ref{guar}),
for a much wider variety of noise conditions than the PP method.\\

\noindent \textbf{Algorithm 1:}\\
For each $b \in \{1,\ldots,B\}$
\begin{enumerate}
  \item Simulate an image, $Y^b$, with the same dimensions as the data, with noise distribution $F$ and no
sources
 \item Record the maximum value of the image, $Y^b_{(N)}$
 \item For k = $1,2,\ldots a$ randomly remove k pixels and record the maximum
remaining pixel value $Y^b_{(N-k)}$. $N-a$ is the smallest number of pixels
for which you want to calculate the p-value, it needs to be more than the number
of pixels that you believe to be sources, but the number of computations increase with $a$
\end{enumerate}

In this way, we build the distribution of the maximum for a given area under the
null. Then we can calculate the appropriate p-values

\begin{equation}
 p(x,A)=P_0(\max (Y_{ij} \in A) \geq x)
\end{equation}

which were previously approximated using Piterbarg's formula and are now estimated
using the simulated data 

\begin{equation}
 p(x,A) \approx \frac{\#\{1\leq i \leq B : Y^i_{\mbox{\tiny{(Area(\it{A}}))}}\geq x\}}{B}
\end{equation}
Then we proceed as in \citet{Pacifico04} to calculate $U$. 

\begin{figure}[h!tp] 
        \centerline{\includegraphics[width=2.5in,angle=270]{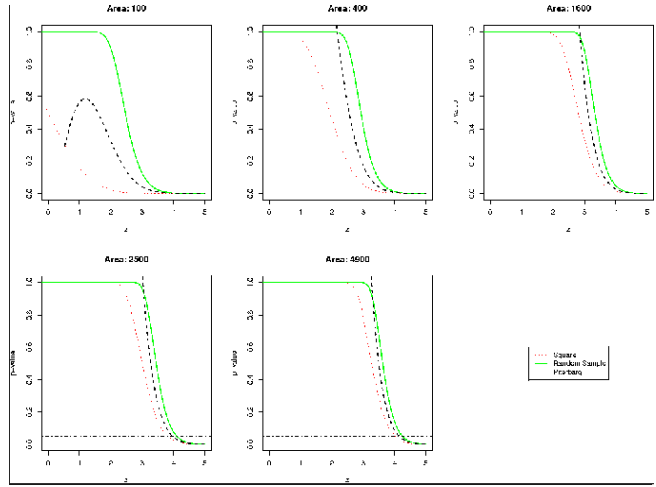}}
        \caption{P-values for a Gaussian Random field that satisfies Equation
\ref{quad} for different sized regions. The dotted black line is the value from
Piterbarg that does not consider any shape information, the solid green line is
calculated using a simulation that randomly samples regions of area 1, the
dotted red line is using the same simulation but the area used is from a square
region. Shape information appears to have a mild effect that gets smaller as the
area considered gets larger. Also Piterbarg exhibits strange behavior for small
areas. The simulation was run 10,000 times.}
        \label{fig:pvals}
\end{figure}

For a Gaussian Random Field that satisfies Equation \ref{quad} we can compare
the Piterbarg formula to our simulated p-values. In Figure \ref{fig:pvals} we
simulate the p-values for different sized areas A, as described above and also
by
replacing Step 3 with removing a square region of area A from the image. We see
that the p-value from Piterbarg, which accounts for size but not shape, tends
to fall in between these two but the differences are small and get smaller as the area gets bigger. 
For astronomical source detection, the random sampling approach is
closest to the data we observe so we use it as our default. 
However, it is worth noting that our simulation technique is easily
adaptable to handle prior knowledge about the shapes of the objects of interest.
In addition, unlike Piterbarg, our simulation procedure gives accurate p-values for any x, not just those in the tail.
However, the biggest benefit
of simulating the p-values is we are not restricted to smooth Gaussian Random
Fields and instead can handle a wider range of noise conditions.

We can simplify the simulations further. We note that $P(x,A)>P(x,A-k)$ 
$\forall k>0$. Since we are dealing with large images,
where the source signal is sparse relative to the background,
the maximum over N pixels
is not going to be substantially different than the maximum over N-k pixels when
$N>>k$. Thus if we use the following algorithm we will get a comparable confidence
superset directly and with less computation.\\

\noindent \textbf{Algorithm 2:}
\begin{enumerate}
 \item For each $b \in \{1,\ldots,B\}$, simulate an image, $Y^b$, with the same dimensions as the data, with noise distribution F and no
sources, then record its maximum
 \item Calculate $r_{1-\alpha}$, the $1-\alpha$ percentile of the maxima
 \item Using the original data image, calculate the level set $U= (L_{r_{1-\alpha}})^{C}$
\end{enumerate}
Then $U$ is a $1-\alpha$ confidence superset for the null. Using the
second algorithm is faster because we avoid the computation of the aB
maxima at the cost of having a bigger confidence superset $U$. However
as long as $N>>k$, which will be the case for telescope images where
sources are relatively sparse on the sky, the size difference is
negligible. The second algorithm also allows us to calculate $U$
directly without having to search over sets or approximate any p-values.

These simulations are only meant to approximate a confidence superset for a given null distribution -- 
it is not meant to simulate the complex systems in the real data.
Therefore we do not need to use prior knowledge or make assumptions about the underlying Astronomy as in the simulations commonly used by astronomers for source detection.
Our simulated confidence superset $U$ can be calculated for any noise
condition from which we can sample numerically.
Furthermore we can also simulate what
happens if we apply filters to the data as is commonly performed in astronomy.
Filters can often be used to enhance sources and remove background
contamination, making detection easier. For instance if we know our data have a
Poisson noise distribution, we can simulate the Poisson noise image, apply the
filter, and then calculate the maxima of the filtered image. We then can
calculate the confidence superset for filtered Poisson data. This combination of
filtering and false cluster proportion control will give us a very powerful tool
 for source detection.

\section{Using Multi-scale Derivatives to Enhance Detection Power}

We have seen with the Chandra data that FCP, using a Z-test and the PP
method, does not achieve power comparable to the methods in \citet{1msec}[GI].  In
the previous section, we outlined a simulation-based approach which
replaces the PP method, allowing for more general test statistics that
do not necessarily have to be Gaussian.  In this section, we describe
a new technique which directly address two problems contributing to
the lack of power seen in Section 4.  The first problem is that the
Z-test assumes a constant background mean.  While this is often a reasonable
assumption, there may be locations in the image where
this does not hold; for example at the edge of exposures where certain
pixels may have been exposed longer than others.  In general,
telescope images can have non-constant backgrounds that vary
spatially, in which case the constant background mean is clearly
violated.  The second problem is that rules based on the magnitude of
a pixel, like the Z-test, are not necessarily the best at
discriminating sources from background.  To address both of these
problems we have developed a Multi-scale Derivative approach which
uses local information to identify regions that stand out from the background as sources.

\begin{figure}[h!tp] 
        \centerline{\includegraphics[width=.8\textwidth]{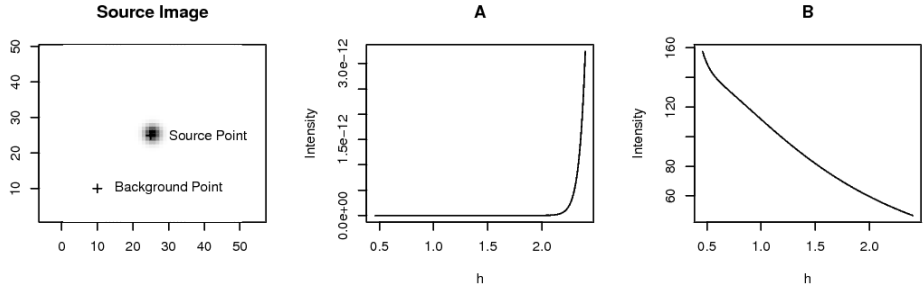}}
        \caption{Left:A toy source with a background and a source point
highlighted. A: The intensity of the background point after smoothing with
different values of the smoothing parameter h. The intensity stays constant
around zero until the source begins to get smoothed into the background, at
which point we see a small increase in the intensity. B: The intensity of a
point in the source after smoothing with different values of the smoothing
parameter h. The intensity drops sharply for small amounts of smoothing and
continues to decline for larger values of h. Thus estimates of the derivative of
the source point with respect to h should give large negative values whereas the
background point will not. }
        \label{fig:msc}
\end{figure}

The idea behind Multi-scale Derivatives is to apply a Gaussian filter
to the data using a range of bandwidths and examine the derivative of
the filtered image with respect to the bandwidth. To get an idea of
why this works, we examine a toy example as shown in Figure
\ref{fig:msc}.  Suppose we are at a location with no signal. At small
bandwidths, our smooth estimate of the null location remains low, and
this will stay roughly the same until we reach a bandwidth that is big
enough to hit a source, at which point our smoothed estimate will jump
up as the source gets smoothed into the null space. The bandwidth at
which this jump occurs tells us how close we are to a
source. Alternatively, if we are at a source, we will have a high
value for our smooth estimate if our bandwidth is small and as we
increase the bandwidth, our estimate will get lower and lower as null
areas are smoothed into the source.

If we compare the derivative of our smoother with respect to the
bandwidth, sources will stand out by their large negative derivatives.
The smoothing operation can be calculated quickly by performing the
convolution in Fourier space, and we can use the same trick to
calculate the required derivative of the smoother.  The derivative is
calculated by convolving the data with a filter of the form
\begin{equation}
F_h(u,v)\propto (\frac{d(u,v)^2}{h^3} -\frac{2}{h}) *\phi(u,v|h) 
\end{equation}
where $d(u,v)$ is the distance between the origin and point
$(u,v)$. $\phi$ is a symmetric Gaussian kernel with bandwidth $h$.  We
then choose a value for $h$ and convolve the filter with the image to
get an image which estimates the derivative for each pixel at the
scale $h$. We do this for a few values of $h$ which are selected to
cover the range of source sizes. The Multi-scale Derivative image, $M$,
is then created at each pixel by selecting the minimum derivative at
that pixel over all the scales.
\begin{equation}
 M(u,v) = \min_h F_h(u,v)
\end{equation}

\begin{figure}[h!tp] 
        \centerline{\includegraphics[scale=.5]{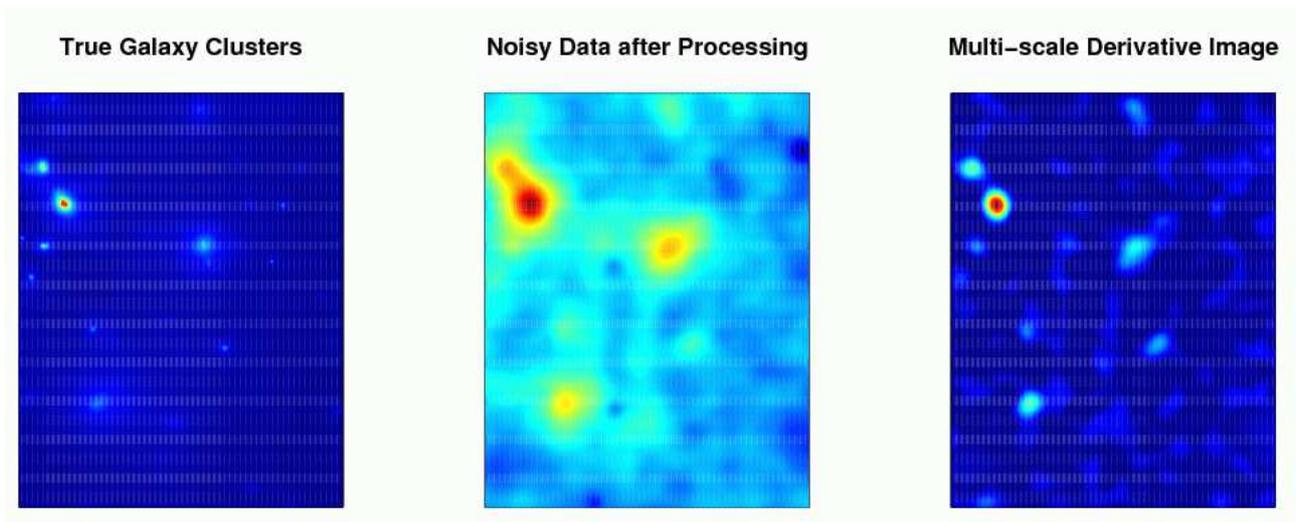}}
        \caption{Left: The galaxy clusters we want to detect. In
          practice these will be obscured by confounding radio signals
          Middle: The Wiener filtered image which combines information
          from three different frequencies in an attempt to pick out
          the signal from the galaxy clusters while eliminating the
          confounding signals. This filter has introduced a
          non-constant background to the image Right: The Multi-scale
          Derivative image calculated from the Wiener filtered image
          recovers most of the sources we are trying to detect while
          getting rid of most of the background variation. Data courtesy of \citet{sehgal} }
        \label{fig:mscSZ}
\end{figure}

 As a result, we should see high negative values at the location of
sources in $M$.  The Multi-scale Derivative image $M$ measures the
peakedness in the data caused by sources in cases where the background
is smoothly varying and will enhance sources in images where there is
a flat background. As a demonstration, in Figure \ref{fig:mscSZ} we
show a small dataset due to \citet{sehgal} simulated to look like data from the Atacama
Cosmology Telescope \citep{arthur}, a ground-based radio telescope
used to detect large galaxy clusters.
The clusters we wish to detect
are visible as bright spots in the simulated image on the left. In
practice, we cannot observe these clusters directly because they are
obscured by several other stronger radio sources. To isolate the
clusters, \citeauthor{sehgal} use a Wiener filter, which combines data
from three different radio frequencies, as seen in the middle
image. Our goal is to detect the galaxy clusters from the filtered
image. The brightest sources are visible in the Wiener filtered image, but
the background is spatially varying and is much brighter in certain
areas than in others. Any detection algorithm based on the magnitude
of pixels will not perform well on this image. Instead we calculate
the Multi-scale Derivative image from the filtered image as seen in
the right panel of Figure \ref{fig:mscSZ}. The Multi-scale Derivative
image suppresses the background and highlights the sources, which will
make source detection much easier. In this example, the Multi-scale
Derivative is used as a filter which enhances sources in the image
while suppressing background. Alternatively, it can be viewed as a
test-statistic which incorporates local information to test for peaks
in the data.  Multi-scale Derivatives can be combined with the
simulation procedure described in Section 5 to create a catalog with
False Cluster Proportion control, the MSFCP procedure. We expect that the catalog created
this way, which is specifically designed to draw out sources, will
give us more detection power than using a simple Z-test statistic.

\section{MSFCP on Chandra Data}
We expect that using the Multi-scale Derivative filter on the CDFS
image will enhance the clusters, making them stand out more from the
background. After performing the smooth-then-root transformation as
described previously, we compute the Multi-scale Derivative image. We
then ran FCP on the Multi-scale Derivative image $M$, using a confidence
superset that was obtained using Algorithm 2 as outlined in Section 5.
We will refer to this procedure as the MSFCP procedure.  Using MSFCP with the same parameters as before
we detect 24 of the 26 sources detected in the main catalog. We do not
make any detections outside of the catalog. Since the detections in
the catalog have been replicated in follow-up observations this
suggests that we are not making any false detections while still
detecting almost all of the real objects.

\begin{figure}[h!tp] 
        \centerline{\includegraphics[scale=.35,angle=90]{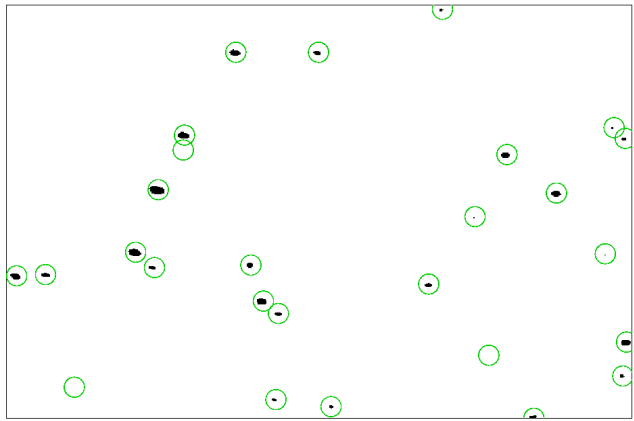}}
        \caption{CDFS data after applying MSFCP procedure. Black indicates sources detected by FCP
and the circles are detections in both the Main Catalog and the Secondary
Catalogs from GI}
        \label{fig:mscfcp}
\end{figure}
Without having to simulate images made to look like the telescope
images, make any assumptions about the underlying astronomy, and without conducting follow-up
observations, we can still make the statement that with probability .95 less than
10\% of our detections will be false. The other algorithms cannot make such an
explicit guarantee of a pure sample.

The detection strategy used in \citet{1msec}, which we refer to as GI, 
is designed for a small patch of sky with the ability to follow-up all potential
detections. 
In this scenario, it makes sense to use a detection scheme that
casts a wide net.
Because we can go back and verify or reject each potential
detection we will tolerate a larger number of spurious sources initially in
exchange for the ability to detect some fainter sources. 
This wide net strategy is unfair to our MSFCP approach which was run to
keep the false detection rate less than 10\% without the benefit of additional data.
To mimic this wide net
approach we can relax our parameters. If we allow for the
proportion of false sources to grow to 20\% we detect 26 sources, two of which
are new detections outside of the main catalog. 
We do not have the resources to follow up both of these new detections,
however one of the new detections is close to two detections in the published catalog.
This allows us to look at the follow-up optical observation for that area to verify our new detection.

\begin{figure}[h!tb] 
        \centerline{\includegraphics[width=5in]{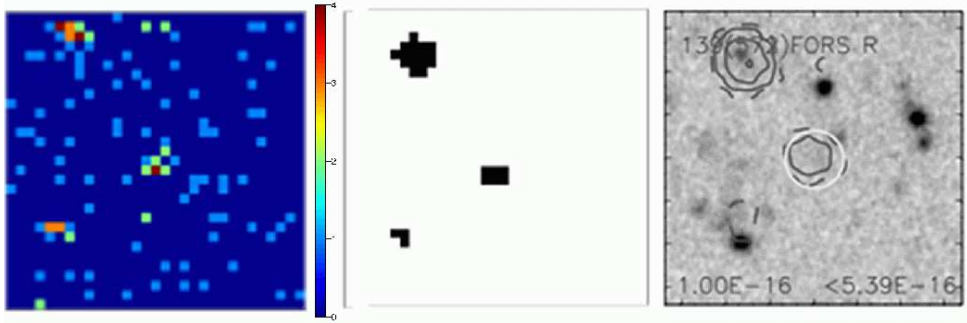}}
        \caption{Left: Raw photon counts from the telescope.
	Center: Black regions are the three sources detected by MSFCP using a tolerance of 20\%. 
	The detection in the middle and upper left are also detected and verified in GI.
	Left: The optical counterpart observation for the source in the center from \citet{1msec}.
	The contour lines indicate X-ray activity at 3,5,10, and 20 $\sigma$ above the local background.
	We can clearly see a source that overlaps with our detection in the lower left, 
	suggesting it is not a spurious detection.}
        \label{fig:newDetection}
\end{figure}

In Figure \ref{fig:newDetection}, we show the original X-ray photon
counts for the region.  Visually we see three candidates and using a
False Cluster Proportion tolerance of 20\% we detect all three.  In
the final panel of Figure \ref{fig:newDetection} we see the optical
follow up which clearly shows an object at the location of our new
detection.  Thus we are confident that this new detection is in fact a
real source that was missed by the analysis in GI.  Our procedure has
not only captured almost all of the detections from the previous
analysis but with rigorous error control, it has also found a new
source that was missed by the astronomers.

\begin{center}
\begin{tabular}{|l|l|l|l|}
\hline
\multicolumn{4}{|c|}{Comparison of Detection Procedures} \\
\hline
Method & False Cluster Proportion &\% of Main Catalog & \# of New Verified  \\ 
 &  Tolerance ($c$) & in GI recovered & Detections outside of GI \\ \hline
  PP & 10\% & 65.3\% & 0 \\
  MSFCP & 10\% & 92.3\% & 0 \\
  MSFCP & 20\% & 92.3\% & 1 \\ \hline

\end{tabular}
\end{center}

We do not have corresponding optical data for our other detection so
we are unable to verify whether it is a real detection or not.  If it
is real we would have no spurious detections while making two new
detections.  Even if it is not real we would still have an error rate
of 3.8\% which is well under the 20\% we are willing to accept.  We
expect the proportion of false sources one is willing to tolerate to
be a parameter that astronomers will have to determine to suit the
problem at hand.  For instance, the ``wide-net'' approach works
adequately for small data sets, but newer telescopes, like LSST, will
collect such a large amount of data that following up each potential
detection is impractical. In this scenario having a detection
procedure that controls for the proportion of false detection
automatically in one pass will be vitally important.

\section{Beyond Astronomy: An illustration for other types of objects}

To illustrate the reach of our method, we apply it to data from a fMRI
experiment.  We do not aim here to provide a definitive analysis of
these data, but rather to show how our method can be adapted beyond
astronomical images.  We will report on a full analysis of these data
in another paper.

In a functional Magnetic Resonance Imaging experiment, a participant
is placed in the Magnetic Resonance scanner and asked to perform a
carefully arranged sequence of behavioral tasks while
three-dimensional brain images are acquired at regular intervals.
Concentrated neural activity produced by performing the tasks induces
detectable but subtle changes in the images due to a blood-flow
response in the brain.  A common form of analysis for fMRI data
considers the time series of measurements at each location in the
image and computes a statistic for that location that is designed to
detect task-related changes in the signal.  ``Hot spots'' in images of
these statistics roughly correspond to areas of activity in the brain.
See \citet{genovese2000} for more detail on this process.

Here we consider a simple experiment in which the participant is asked
to visually fixate on a spot at the center of the visual field while
two complementary sets of annular rings flash alternately at a fixed
frequency.  The images were acquired using a new technique 
that provides much higher spatial
resolution than is typical for fMRI studies without sacrificing
temporal resolution.  The three-dimensional images are then projected
down to the two-dimensional surface of the brain.  We thank Drs.~David
Heeger and Eli Merriam at NYU for the use of these data.

At each location in the primary area responsible for visual processing
(called V1), the fMRI signal should vary roughly like a sinusoid, as
the corresponding location in the visual field turns on and off.
Locations that respond to the first set of rings should have
sinusoidal signals $180^\circ$ out of phase to locations that respond
to the second set of rings. 

Our goal is to identify the regions in V1 that activate in response to
each set of stimuli.  This will appear as a series of bands across the
visual cortex with alternating bands out of phase with each other.  As
in the astronomical problems, we are interested in the active regions
themselves rather than the individual pixels, so we prefer an
inferential method that can give us control over regions as opposed to
pixels.

To detect the periodicity at each location, we use a variant of
Fisher's F-test \citep{brockwell} for each time series to determine if
that location is responsive to the stimulus.  For each location, we
also have the response phase which, for locations that respond to the
stimulus, we expect to cluster corresponding to the differing sets of
rings. This leaves us with three basic types of locations which we
want to differentiate, null locations for which there is no response
to the stimulus, locations that respond to the first set of rings and
those that respond to the second set of rings.  We run a simple
classifier on the phases to partition the locations into two phase classes which correspond to the two different stimulus rings. We then want to conduct a test that will group
locations into clusters corresponding to the phases with a
probabilistic guarantee on the rate of false clusters. To do this we
need to adapt our definition of a cluster to both deal with having
multiple classes of clusters and also to dealing with points in space
as opposed to pixels on a regular grid. For multiple classes of
clusters we first have to classify each location into a class and then
we modify our definition of a cluster such that a cluster must be made
up of locations all belonging to the same class.

To handle the points in space, we define a graph on the points.  We
call two locations connected at level $t$ if they are both of the same
class, have p-values from the F-test less that $t$ and there is a path
between the two locations in which each node is also of the same
class, has p-values less than $t$ and no edge has euclidean distance
greater than $d$. Then the $i$th connected component at level $t$,
$C_{i,t}$, is the largest set of locations that are all connected.

\begin{figure}[ht] 
        \centerline{\includegraphics[width=3in]{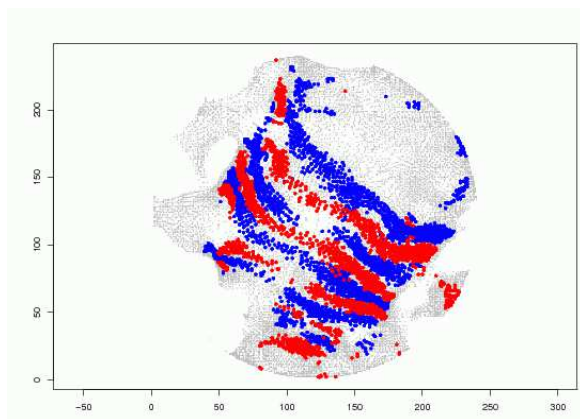}}
        \caption{Gray dots represent every location recorded
while blue and red dots indicate locations in clusters that were declared active.
The two colors denote the two different phases, one corresponding to the rings
and the other to the anti-rings. We clearly see the banded activation patterns
expected in the data.}
        \label{fig:fmri}
\end{figure}

We also need to derive a confidence superset $U$. Recall, $U$ is a
$1-\alpha$ confidence superset for $S_0$ if $P(U \supseteq S_0) \geq 1
- \alpha$. Our null has already been determined by the F-test and for
each location we have the p-value of that test. A simple way to calculate $U$ is to find the $(1-\alpha)$th percentile of the
distribution of the maximum of the p-values, which will have a uniform
distribution under the null. All locations with p-value greater than
$1-(1-\alpha)^{\frac{1}{n}}$ belong to the set $U$, where $n$ is the total
number of locations. This will provide a conservative confidence
superset 
and future work will use local information to obtain a smaller confidence superset.  Now we can proceed
similar to the previous cases. For different values of $t$ we examine
the connected components and if the fraction of location in $C_{i,t}$
that are also in $U$ is greater than $\epsilon$ we declare the cluster
$C_{i,t}$ false. We proceed analogous to the case of a pixelized
image, searching for the value of t such that the rate of false
regions is bounded appropriately with high probability.  Figure
\ref{fig:fmri} shows the result of an experiment in which we set our
detection limit so that with probability .95 less than 10\% of the
clusters detected are false. We can clearly see the bands that we are
hoping to detect as well as some active areas along the edges which
may be artifacts of the projection from three dimensions to the
surface of the brain. This is a demonstration that False Cluster
Proportion concepts can be extended to various situations where
controlling the rate of falsely detected regions is desirable to detect
structure. Here we have shown that we can get the same false cluster
proportion control when there is more than one class of objects and
also when our data do not lie a regular grid.

\section{Conclusion}
We have extended the False Cluster Proportion Multiple Testing Procedure so it
can now be applied to a wide variety of problem in Astronomy as well as other
fields. By moving away from theoretical approximations, like Piterbarg's
approximation, we can now calculate confidence supersets for a large variety of
noise conditions. This allows us to create source catalogs with a
probabilistic guarantee that they are not overly polluted with false detections
without having to make any difficult to check assumptions about the data or the
science behind the data. The Multi-scale Derivative procedure enhances the types of
sources we typically see in a telescope image and we have shown that they can
also enhance the power of False Cluster Proportion source detection algorithms.
This is evidenced by our analysis of the CDFS data. The catalog
published in GI uses two different algorithms and then has to follow up each
detection, whereas we run MSFCP and can make the statement that with
probability .95 less than 10\% of our detections are false, and still get
virtually the same catalog. 
When we allow 20\% of our detections to be false we make two new detections,
one of which we have verified is a real source that was missed in the original analysis.
Our procedure provides rigorous error control,
which is lacking in the current techniques used by astronomers. 
At the same time we have demonstrated that the detection power is comparable and
in fact we have detected sources that they have missed.
Controlling false detections without the
need for expensive follow up observations will be critical as the next generation
of telescopes will provide a deluge of data that will be impossible to process
manually.

We have also shown that these ideas can be generalized to other types of
problems where we are dealing with points spread along a plane instead of a
regular grid of pixels in an image. We can further modify our definition of
cluster to differentiate between different types of objects. We can then control
the proportion of false clusters, allowing for different classes of clusters. In the
future, we believe we can further extend our framework so that we can apply
False Cluster Proportion techniques on a wide variety of statistical clustering
problems, in which we are trying to detect clusters of data without making false
detections.

{\scriptsize
  \bibliography{jasa}

\begin{thebibliography}{43}
\newcommand{\enquote}[1]{``#1''}
\expandafter\ifx\csname natexlab\endcsname\relax\def\natexlab#1{#1}\fi

\bibitem[{{Anscombe}(1948)}]{anscombe}
{Anscombe}, F. (1948), \enquote{The transformation of Poisson, binomial and
  negative-binomial data,} \textit{Biometrika}, 35, 246--254.

\bibitem[{Benjamini and Hochberg(1995)}]{bh}
Benjamini, Y. and Hochberg, Y. (1995), \enquote{Controlling the False Discovery
  Rate: a Practical and Powerful Approach to Multiple Testing,} \textit{Journal
  of the Royal Statistic Society}, B, 289--300.

\bibitem[{Benjamini et~al.(2006)Benjamini, Krieger, and Yekutieli}]{stepup}
Benjamini, Y., Krieger, A.~M., and Yekutieli, D. (2006), \enquote{Adaptive
  Linear Step-Up Procedures that control the False Discovery Rate,}
  \textit{Biometrika}, 93, 491--507.

\bibitem[{{Bertin} and {Arnouts}(1996)}]{sext}
{Bertin}, E. and {Arnouts}, S. (1996), \enquote{{SExtractor: Software for
  source extraction.}} \textit{\aaps}, 117, 393--404.

\bibitem[{Brockwell and Davis(1998)}]{brockwell}
Brockwell, P.~J. and Davis, R.~A. (1998), \textit{Time Series: Theory and
  Methods (Springer Series in Statistics)}, Springer.

\bibitem[{Buckner(1998)}]{neuro}
Buckner, R.~L. (1998), \enquote{Event-related f{MRI} and the hemodynamic
  response,} \textit{Human Brain Mapping}, 6, 373--377.

\bibitem[{{Carvalho} et~al.(2009){Carvalho}, {Rocha}, and {Hobson}}]{fastBayes}
{Carvalho}, P., {Rocha}, G., and {Hobson}, M.~P. (2009), \enquote{{A fast
  Bayesian approach to discrete object detection in astronomical data sets -
  PowellSnakes I},} \textit{\mnras}, 393, 681--702.

\bibitem[{{Cash}(1979)}]{poissModel}
{Cash}, W. (1979), \enquote{{Parameter estimation in astronomy through
  application of the likelihood ratio},} \textit{\apj}, 228, 939--947.

\bibitem[{Damiani et~al.(1997)Damiani, Maggio, Micela, and Sciortino}]{Damiani}
Damiani, F., Maggio, A., Micela, G., and Sciortino, S. (1997), \enquote{A
  Method Based on Wavelet Transforms for Source Detection in Photon-counting
  Detector Images. I. Theory and General Properties,} \textit{The Astrophysical
  Journal}, 483, 350--369.

\bibitem[{Freeman et~al.(2002)Freeman, Kashyap, Rosner, and Lamb}]{peter}
Freeman, P.~E., Kashyap, V., Rosner, R., and Lamb, D.~Q. (2002), \enquote{A
  Wavelet-Based Algorithm for the Spatial Analysis of Poisson Data,}
  \textit{The Astrophysical Journal Supplement Series}, 138, 185--218.

\bibitem[{Genovese(2000)}]{genovese2000}
Genovese, C.~R. (2000), \enquote{A Bayesian Time-Course Model for Functional
  Magnetic Resonance Imaging Data,} \textit{Journal of the American Statistical
  Association}, 95, 691--703.

\bibitem[{{Genovese} et~al.(2004){Genovese}, {Miller}, {Nichol}, {Arjunwadkar},
  and {Wasserman}}]{npicmb}
{Genovese}, C.~R., {Miller}, C.~J., {Nichol}, R.~C., {Arjunwadkar}, M., and
  {Wasserman}, L. (2004), \enquote{Nonparametric Inference for the Cosmic
  Microwave Background,} \textit{Statistical Science}, 19, 308--321.

\bibitem[{Giacconi et~al.(2002)Giacconi, Zirm, Wang, Rosati, Nonino, Tozzi,
  Gilli, Mainieri, Hasinger, Kewley, Bergeron, Borgani, Gilmozzi, Grogin,
  Koekemoer, Schreier, Zheng, and Norman}]{1msec}
Giacconi, R., Zirm, A., Wang, J., Rosati, P., Nonino, M., Tozzi, P., Gilli, R.,
  Mainieri, V., Hasinger, G., Kewley, L., Bergeron, J., Borgani, S., Gilmozzi,
  R., Grogin, N., Koekemoer, A., Schreier, E., Zheng, W., and Norman, C.
  (2002), \enquote{Chandra Deep Field South: The 1 Ms Catalog,} \textit{The
  Astrophysical Journal Supplement Series}, 139, 369--410.

\bibitem[{Gonzalez-Nuevo et~al.(2006)Gonzalez-Nuevo, Argueso, Lopez-Caniego,
  Toffolatti, Sanz, Vielva, and Herranz}]{gonzaleznuevo-2006-369}
Gonzalez-Nuevo, J., Argueso, F., Lopez-Caniego, M., Toffolatti, L., Sanz,
  J.~L., Vielva, P., and Herranz, D. (2006), \enquote{The Mexican Hat Wavelet
  Family. Application to point source detection in CMB maps,}
  \textit{MON.NOT.ROY.ASTRON.SOC.}, 369, 1603.

\bibitem[{{Guglielmetti} et~al.(2009){Guglielmetti}, {Fischer}, and
  {Dose}}]{gug}
{Guglielmetti}, F., {Fischer}, R., and {Dose}, V. (2009),
  \enquote{{Background-source separation in astronomical images with Bayesian
  probability theory - I. The method},} \textit{\mnras}, 396, 165--190.

\bibitem[{Heller et~al.(2006)Heller, Stanley, Yekutieli, Rubin, and
  Benjamini}]{cba}
Heller, R., Stanley, D., Yekutieli, D., Rubin, N., and Benjamini, Y. (2006),
  \enquote{Cluster-based analysis of FMRI data.} \textit{Neuroimage}, 33,
  599--608.

\bibitem[{{Herschel}(1786)}]{catalogue-of-nebulae}
{Herschel}, W. (1786), \enquote{{Catalogue of One Thousand New Nebulae and
  Clusters of Stars. By William Herschel, LL.D. F. R. S.}} \textit{Royal
  Society of London Philosophical Transactions Series I}, 76, 457--499.

\bibitem[{{Hobson} and {McLachlan}(2003)}]{2003MNRAS.338..765H}
{Hobson}, M.~P. and {McLachlan}, C. (2003), \enquote{{A Bayesian approach to
  discrete object detection in astronomical data sets},} \textit{\mnras}, 338,
  765--784.

\bibitem[{{Hopkins} et~al.(2002){Hopkins}, {Miller}, {Connolly}, {Genovese},
  {Nichol}, and {Wasserman}}]{2002AJ....123.1086H}
{Hopkins}, A.~M., {Miller}, C.~J., {Connolly}, A.~J., {Genovese}, C., {Nichol},
  R.~C., and {Wasserman}, L. (2002), \enquote{{A New Source Detection Algorithm
  Using the False-Discovery Rate},} \textit{\aj}, 123, 1086--1094.

\bibitem[{Kosowsky(2003)}]{arthur}
Kosowsky, A. (2003), \enquote{The Atacama Cosmology Telescope,} \textit{New
  Astronomy Reviews}, 47, 939.

\bibitem[{{Loredo}(2007)}]{loredo}
{Loredo}, T.~J. (2007), \enquote{{Analyzing Data from Astronomical Surveys:
  Issues and Directions},} in \textit{Statistical Challenges in Modern
  Astronomy IV}, eds. {Babu}, G.~J. and {Feigelson}, E.~D., vol. 371 of
  \textit{Astronomical Society of the Pacific Conference Series}, pp. 121--+.

\bibitem[{{Markevitch} et~al.(2004){Markevitch}, {Gonzalez}, {Clowe},
  {Vikhlinin}, {Forman}, {Jones}, {Murray}, and {Tucker}}]{bullet}
{Markevitch}, M., {Gonzalez}, A.~H., {Clowe}, D., {Vikhlinin}, A., {Forman},
  W., {Jones}, C., {Murray}, S., and {Tucker}, W. (2004), \enquote{{Direct
  Constraints on the Dark Matter Self-Interaction Cross Section from the
  Merging Galaxy Cluster 1E 0657-56},} \textit{\apj}, 606, 819--824.

\bibitem[{Meinshausen and Rice(2006)}]{rice}
Meinshausen, N. and Rice, J. (2006), \enquote{Estimating the proportion of
  false null hypotheses among a large number of independently tested
  hypotheses,} \textit{Ann. Statist}, 34, 373--393.

\bibitem[{{Melin} et~al.(2006){Melin}, {Bartlett}, and {Delabrouille}}]{melin}
{Melin}, J.-B., {Bartlett}, J.~G., and {Delabrouille}, J. (2006),
  \enquote{{Catalog extraction in SZ cluster surveys: a matched filter
  approach},} \textit{\aap}, 459, 341--352.

\bibitem[{Perone~Pacifico et~al.(2004)Perone~Pacifico, Genovese, Verdinelli,
  and Wasserman}]{Pacifico04}
Perone~Pacifico, M., Genovese, C., Verdinelli, I., and Wasserman, L. (2004),
  \enquote{False Discovery Control for Random Fields,} \textit{Journal of the
  American Statistical Association}, 99, 1002--1014.

\bibitem[{Piterbarg(1996)}]{pite}
Piterbarg, V. (1996), \textit{Asymptotic Methods in the theory of Gaussian
  processes and fields}, American Mathematical Society.

\bibitem[{Richards and Jia(1999)}]{remote}
Richards, J.~A. and Jia, X. (1999), \textit{Remote Sensing Digital Image
  Analysis: An Introduction}, Secaucus, NJ, USA: Springer-Verlag New York, Inc.

\bibitem[{{Richards} et~al.(2009){Richards}, {Freeman}, {Lee}, and
  {Schafer}}]{richards}
{Richards}, J.~W., {Freeman}, P.~E., {Lee}, A.~B., and {Schafer}, C.~M. (2009),
  \enquote{{Exploiting Low-Dimensional Structure in Astronomical Spectra},}
  \textit{\apj}, 691, 32--42.

\bibitem[{{Savage} and {Oliver}(2007)}]{2007ApJ...661.1339S}
{Savage}, R.~S. and {Oliver}, S. (2007), \enquote{{Bayesian Methods of
  Astronomical Source Extraction},} \textit{\apj}, 661, 1339--1346.

\bibitem[{{Sehgal} et~al.(2007){Sehgal}, {Trac}, {Huffenberger}, and
  {Bode}}]{sehgal}
{Sehgal}, N., {Trac}, H., {Huffenberger}, K., and {Bode}, P. (2007),
  \enquote{{Microwave Sky Simulations and Projections for Galaxy Cluster
  Detection with the Atacama Cosmology Telescope},} \textit{The Astrophysical
  Journal}, 664, 149--161.

\bibitem[{Shim et~al.(2006)Shim, Im, Pak, Choi, Fadda, Helou, and
  Storrie-Lombardi}]{spitz}
Shim, H., Im, M., Pak, S., Choi, P., Fadda, D., Helou, G., and
  Storrie-Lombardi, L. (2006), \enquote{Deep u*- and g-Band Imaging of the
  Spitzer Space Telescope First Look Survey Field: Observations and Source
  Catalogs,} \textit{The Astrophysical Journal Supplement Series}, 164,
  435--449.

\bibitem[{Storey(2002)}]{storey}
Storey, J.~D. (2002), \enquote{A Direct Approach to False Discovery Rates,}
  \textit{Journal of the Royal Statistical Society. Series B (Statistical
  Methodology)}, 64, 479--498.

\bibitem[{Strong(2003)}]{strong}
Strong, A.~W. (2003), \enquote{Maximum Entropy imaging with INTEGRAL/SPI data,}
  \textit{A\&A}, 411, L127--L129.

\bibitem[{Sun and Cai(2007)}]{suncai}
Sun, W. and Cai, T.~T. (2007), \enquote{Oracle and Adaptive Compound Decision
  Rules for False Discovery Rate Control,} \textit{Journal of the American
  Statistical Association}, 102, 901--912.

\bibitem[{Tyson and the LSST~Collaboration(2002)}]{lsst}
Tyson, J.~A. and the LSST~Collaboration (2002), \enquote{Large Synoptic Survey
  Telescope: Overview,} \textit{PROC.SPIE INT.SOC.OPT.ENG.}, 4836, 10.

\bibitem[{Vale and White(2006)}]{vale}
Vale, C. and White, M. (2006), \enquote{Finding Clusters in SZ Surveys,}
  \textit{NEW ASTRON.}, 11, 207.

\bibitem[{{Valtchanov} et~al.(2001){Valtchanov}, {Pierre}, and
  {Gastaud}}]{valt}
{Valtchanov}, I., {Pierre}, M., and {Gastaud}, R. (2001), \enquote{{Comparison
  of source detection procedures for XMM-Newton images},} \textit{\aap}, 370,
  689--706.

\bibitem[{{van Dyk} et~al.(2009){van Dyk}, DeGennaro, Stein, Jefferys, and {von
  Hippel}}]{vandyk-2009-3}
{van Dyk}, D.~A., DeGennaro, S., Stein, N., Jefferys, W.~H., and {von Hippel},
  T. (2009), \enquote{Statistical analysis of stellar evolution,}
  \textit{ANNALS OF APPLIED STATISTICS}, 3, 117.

\bibitem[{{Vikhlinin} et~al.(1995){Vikhlinin}, {Forman}, {Jones}, and
  {Murray}}]{vik}
{Vikhlinin}, A., {Forman}, W., {Jones}, C., and {Murray}, S. (1995),
  \enquote{{Matched Filter Source Detection Applied to the ROSAT PSPC and the
  Determination of the Number-Flux Relation},} \textit{ApJ}, 451, 542--+.

\bibitem[{{Weisskopf} et~al.(2000){Weisskopf}, {Tananbaum}, {van Speybroeck},
  and {O'dell}}]{chandra}
{Weisskopf}, M., {Tananbaum}, H., {van Speybroeck}, L., and {O'dell}, S.
  (2000), \enquote{{Chandra X-Ray Observatory: Overview, X-Ray Optics,
  Instruments, and Missions},} \textit{Proc. SPIE}, 4012.

\bibitem[{Worsley et~al.(2002)Worsley, Liao, Aston, Petre, Duncan, Morales, and
  Evans}]{wor02}
Worsley, K., Liao, J., Aston, C., Petre, V., Duncan, G., Morales, F., and
  Evans, A. (2002), \enquote{A General Statistical Analysis for fMRI data,}
  \textit{NeuroImage}, 15, 1--15.

\bibitem[{Worsley et~al.(1996)Worsley, Marrett, Vandal, Friston, and
  Evans}]{wor96}
Worsley, K., Marrett, S., Vandal, A., Friston, K., and Evans, A. (1996),
  \enquote{A Unified Statistical Approach for Determining Significant
  Statistical Signals in Imagea of Cerebreal Activation,} \textit{Human Brain
  Mapping}, 4, 58--73.

\bibitem[{York et~al.(2000)York, Adelman, Anderson, Jr., Anderson, Annis,
  Bahcall, Bakken, Barkhouser, Bastian, Berman, Boroski, Bracker, Briegel,
  Briggs, Brinkmann, Brunner, Burles, Carey, Carr, Castander, Chen, Colestock,
  Connolly, Crocker, Csabai, Czarapata, Davis, Doi, Dombeck, Eisenstein,
  Ellman, Elms, Evans, Fan, Federwitz, Fiscelli, Friedman, Frieman, Fukugita,
  Gillespie, Gunn, Gurbani, de~Haas, Haldeman, Harris, Hayes, Heckman,
  Hennessy, Hindsley, Holm, Holmgren, hao Huang, Hull, Husby, Ichikawa,
  Ichikawa, Ivezic, Kent, Kim, Kinney, Klaene, Kleinman, Kleinman, Knapp,
  Korienek, Kron, Kunszt, Lamb, Lee, Leger, Limmongkol, Lindenmeyer, Long,
  Loomis, Loveday, Lucinio, Lupton, MacKinnon, Mannery, Mantsch, Margon,
  McGehee, McKay, Meiksin, Merelli, Monet, Munn, Narayanan, Nash, Neilsen,
  Neswold, Newberg, Nichol, Nicinski, Nonino, Okada, Okamura, Ostriker, Owen,
  Pauls, Peoples, Peterson, Petravick, Pier, Pope, Pordes, Prosapio,
  Rechenmacher, Quinn, Richards, Richmond, Rivetta, Rockosi, Ruthmansdorfer,
  Sandford, Schlegel, Schneider, Sekiguchi, Sergey, Shimasaku, Siegmund, Smee,
  Smith, Snedden, Stone, Stoughton, Strauss, Stubbs, SubbaRao, Szalay, Szapudi,
  Szokoly, Thakar, Tremonti, Tucker, Uomoto, Berk, Vogeley, Waddell, i~Wang,
  Watanabe, Weinberg, Yanny, , and Yasuda}]{sdss}
York, D.~G., Adelman, J., Anderson, J.~E., Jr., Anderson, S.~F., Annis, J.,
  Bahcall, N.~A., Bakken, J.~A., Barkhouser, R., Bastian, S., Berman, E.,
  Boroski, W.~N., Bracker, S., Briegel, C., Briggs, J.~W., Brinkmann, J.,
  Brunner, R., Burles, S., Carey, L., Carr, M.~A., Castander, F.~J., Chen, B.,
  Colestock, P.~L., Connolly, A.~J., Crocker, J.~H., Csabai, I., Czarapata,
  P.~C., Davis, J.~E., Doi, M., Dombeck, T., Eisenstein, D., Ellman, N., Elms,
  B.~R., Evans, M.~L., Fan, X., Federwitz, G.~R., Fiscelli, L., Friedman, S.,
  Frieman, J.~A., Fukugita, M., Gillespie, B., Gunn, J.~E., Gurbani, V.~K.,
  de~Haas, E., Haldeman, M., Harris, F.~H., Hayes, J., Heckman, T.~M.,
  Hennessy, G.~S., Hindsley, R.~B., Holm, S., Holmgren, D.~J., hao Huang, C.,
  Hull, C., Husby, D., Ichikawa, S.-I., Ichikawa, T., Ivezic, Z., Kent, S.,
  Kim, R. S.~J., Kinney, E., Klaene, M., Kleinman, A.~N., Kleinman, S., Knapp,
  G.~R., Korienek, J., Kron, R.~G., Kunszt, P.~Z., Lamb, D.~Q., Lee, B., Leger,
  R.~F., Limmongkol, S., Lindenmeyer, C., Long, D.~C., Loomis, C., Loveday, J.,
  Lucinio, R., Lupton, R.~H., MacKinnon, B., Mannery, E.~J., Mantsch, P.~M.,
  Margon, B., McGehee, P., McKay, T.~A., Meiksin, A., Merelli, A., Monet,
  D.~G., Munn, J.~A., Narayanan, V.~K., Nash, T., Neilsen, E., Neswold, R.,
  Newberg, H.~J., Nichol, R.~C., Nicinski, T., Nonino, M., Okada, N., Okamura,
  S., Ostriker, J.~P., Owen, R., Pauls, A.~G., Peoples, J., Peterson, R.~L.,
  Petravick, D., Pier, J.~R., Pope, A., Pordes, R., Prosapio, A., Rechenmacher,
  R., Quinn, T.~R., Richards, G.~T., Richmond, M.~W., Rivetta, C.~H., Rockosi,
  C.~M., Ruthmansdorfer, K., Sandford, D., Schlegel, D.~J., Schneider, D.~P.,
  Sekiguchi, M., Sergey, G., Shimasaku, K., Siegmund, W.~A., Smee, S., Smith,
  J.~A., Snedden, S., Stone, R., Stoughton, C., Strauss, M.~A., Stubbs, C.,
  SubbaRao, M., Szalay, A.~S., Szapudi, I., Szokoly, G.~P., Thakar, A.~R.,
  Tremonti, C., Tucker, D.~L., Uomoto, A., Berk, D.~V., Vogeley, M.~S.,
  Waddell, P., i~Wang, S., Watanabe, M., Weinberg, D.~H., Yanny, B., , and
  Yasuda, N. (2000), \enquote{The Sloan Digital Sky Survey: Technical Summary,}
  \textit{The Astronomical Journal}, 120, 1579--1587.

\end{thebibliography}
}

\end{document}